\documentclass[aps,pre,twocolumn,superscriptaddress,floatfix]{revtex4-1}

\pdfoutput=1

\usepackage{hyperref}
\usepackage{amsmath,amssymb}
\usepackage{float}
\usepackage{bm}
\usepackage{bbm}
\usepackage{graphicx, float, caption}

\usepackage{color}

\newcommand{\addjg}[1]{\textcolor{black}{#1}}

\usepackage[normalem]{ulem}

\newcommand{\config}[0]{\ensuremath{\mathcal{C}}}

\newcommand{\be}{\begin{equation}}
\newcommand{\ee}{\end{equation}}
\newcommand{\bea}{\begin{eqnarray}}
\newcommand{\eea}{\end{eqnarray}}
\newcommand{\eps}{\varepsilon}
\newcommand{\mR}{\mathcal{R}}

\newlength{\longme}

\begin{document}

\title{Nonequilibrium grand-canonical ensemble built from a physical particle reservoir}

\date{\today}

\author{Jules \surname{Guioth}}
\affiliation{DAMTP, Centre for Mathematical Sciences, University of Cambridge, Wilberforce Road, Cambridge CB3 0WA, UK}

\author{Eric \surname{Bertin}}
\affiliation{Univ.~Grenoble Alpes, CNRS, LIPhy, F-38000 Grenoble, France}
\email{eric.bertin@univ-grenoble-alpes.fr}

\begin{abstract}
We introduce a nonequilibrium grand-canonical ensemble defined by considering the stationary state of a driven system of particles put in contact with a nonequilibrium particle reservoir. At odds with its equilibrium counterpart, or with purely formal constructions of a grand-canonical ensemble, this physically-motivated construction yields 
a grand-canonical distribution that depends on the details of the contact dynamics between the system and the reservoir. For non-interacting driven particles, a grand-canonical chemical potential can still be defined, although this chemical potential now differs from that of the reservoir.
However, in the general case, the usual exponential factor (in the particle number) defining the grand-canonical chemical potential, is replaced by the exponential of a non-linear function of the density, this function being proportional to the volume. This case is illustrated explicitly on a one-dimensional lattice model.
Although a grand-canonical chemical potential can no longer be defined in this case, it is possible for a subclass of contact dynamics to generalize the equilibrium fluctuation-response relation by introducing a small external potential difference between the system and the reservoir.
\end{abstract}


\maketitle 

\section{Introduction}

Ensemble equivalence plays a key role in equilibrium statistical physics \cite{touchette2004introduction,touchette2009large,touchette2015equivalence}, and knowing whether such a concept can be extended to nonequilibrium situations is an important issue in view of building a nonequilibrium thermodynamics for steady-states \cite{Oono1998,hayashi2003thermo,sasa2006steady,komatsu2010stationary,komatsu2015exact}. For instance, it would be valuable to know whether a driven stationary system behaves in the same way when its number of particles is fixed, or when it is allowed to exchange particles with a reservoir, corresponding respectively to the nonequilibrium extensions of the canonical and grand-canonical ensembles.
When attempting to build a grand-canonical ensemble for driven steady-state systems, a first issue may be the ability to define a nonequilibrium chemical potential in a thermodynamically consistent way \cite{sasa2006steady,dickman2014inconsistencies,dickman2014failure}. In particular, the study of phase separation in steady-state driven systems has shown that equilibrium concepts need to be generalized
\cite{dickman2016phase,speck2016stochastic,prymidis2016vapour,solon2018generalized-pre,solon2018generalized-njp,paliwal2018chemical}.
Contrary to the case of temperature, for which the lack of energy conservation out of equilibrium hinders a thermodynamically consistent definition in nonequilibrium steady states \cite{Jou03,Cugliandolo11,Levine07,Bertin04,Martens09}, a notion of nonequilibrium chemical potential based on the conservation of the number of particles has been proposed some time ago \cite{bertin2006def,bertin2007intensive}, and tested in numerical simulations of stochastic lattice gases \cite{pradhan2010nonequilibrium,pradhan2011approximate}.
This approach relies on the assumption that the large deviation function of particle density is additive when the system is split into subsystems \cite{bertin2006def,bertin2007intensive,chatterjee2015zeroth}.
Recently, the validity conditions of this assumption have been clarified, by a careful analysis of the coarse-grained dynamics describing the contact between subsystems \cite{guioth2018large,guioth2019nonequilibrium}.
It has been found in particular that if the coarse-grained contact dynamics satisfies both a factorization property and a macroscopic detailed balance property, the large deviation function is additive and a chemical potential can be defined. However, this chemical potential does not satisfy in general an equation of state \cite{guioth2018large,guioth2019nonequilibrium,guioth2019lack}, meaning that it does not depend only on bulk quantities like the density, but also on the contact dynamics itself, at odds with equilibrium situations. A similar lack of an equation of state has also generically been reported for the mechanical pressure in gases of active particles, unless specific symmetries are satisfied \cite{solon2015nat,solon2015prl,winkler2015virial,takatori2014swim,takatori2015towards,speck2016ideal,speck2016stochastic,joyeux2016pressure,fily2018mechanical}.

Having introduced a proper nonequilibrium framework to define chemical potentials for systems in contact, one can try to define a nonequilibrium grand-canonical ensemble and study whether its properties are equivalent to that of the nonequilibrium canonical (fixed particle number) system.
The theoretical framework allowing for the definition of nonequilibrium chemical potentials consists in considering two systems in contact, in the weak exchange rate limit  \cite{chatterjee2015zeroth,guioth2018large,guioth2019nonequilibrium,guioth2020nonadditive}
(however, note that interesting phenomena also appear for non-vanishing exchange rates \cite{Dickman19uphill}).
While the two systems have previously been assumed to have comparable sizes, it is of interest to discuss the case when one of the systems is much larger than the other and plays the role of a particle reservoir. A natural and important question is then to know whether the standard equilibrium thermodynamic structure of the grand canonical ensemble remains essentially valid, or if a different structure emerges in this case. This is the question we explore in this paper.
We show in particular that the physical grand-canonical ensemble obtained by connecting a system to a nonequilibrium reservoir differs from the formal grand-canonical ensemble build by formally replacing the delta function enforcing the conservation of the number of particles by an exponential factor in the number of particles, thereby introducing a chemical potential as a Lagrange multiplier. While these two ways of building a grand-canonical ensemble are equivalent at equilibrium, they lead to different results out of equilibrium, and the physical implementation of a reservoir is probably more meaningful to describe the grand canonical situation in driven systems.

The paper is organized as follows. In Sec.~\ref{sec:gen:framework}, we briefly review the framework introduced in \cite{guioth2018large,guioth2019nonequilibrium,guioth2020nonadditive} to describe the steady state of driven systems in contact exchanging particles at a vanishing rate.
Then in Sec.~\ref{sec:buildingGC}, we build the nonequilibrium grand-canonical distribution obtained by putting a driven system in contact with a particle reservoir, which can itself be in a nonequilibrium steady-state. We also briefly discuss some of the properties of this generalized grand-canonical ensemble, like the equivalence with the canonical (i.e., fixed particle number) ensemble, and the fluctuation-response relation.
Finally we discuss in Sec.~\ref{sec:examples} two different explicit examples of grand-canonical ensembles, a gas of noninteracting active particles, and mass transport model on a lattice.

\section{General framework for systems in contact}
\label{sec:gen:framework}

\subsection{Two systems in weak contact}

In line with our previous works \cite{guioth2018large,guioth2019nonequilibrium,guioth2019lack,guioth2020nonadditive}, we consider the following general framework of two systems in contact in the weak exchange rate limit, that we call for short weak contact.
Our general set-up consists in two stochastic Markovian systems A and B that exchange particles at a low rate as compared to the characteristic frequency of the internal dynamics of each system. Both systems are subject to driving forces $f_A$ and $f_B$ respectively, that break microscopic detailed balance.
The number of particles, volume and density of system $k=A,B$ are respectively denoted as $N_k$, $V_k$ and $\rho_k=N_k/V_k$. The total number of particles $N_{\rm T} = N_A+N_B$ is fixed.
The microscopic contact dynamics between the two systems is assumed to be orthogonal to the driving force, in the sense of the classification of contacts proposed by Sasa and Tasaki \cite{sasa2006steady}.
As a consequence, the contact dynamics does not depend on the driving forces \cite{pradhan2010nonequilibrium,pradhan2011approximate,guioth2018large,guioth2019nonequilibrium} and satisfies the local detailed balance with respect to the equilibrium distribution.
However, the contact dynamics does not satisfy in general the microscopic detailed balance relation with respect to the steady-state distributions at non-zero drives.

\subsection{Large deviations of particle densities}

We are specifically interested in determining the joint distribution $P(\rho_A,\rho_B)$ of particles densities $\rho_A$ and $\rho_B$.
It has been argued in \cite{guioth2018large,guioth2019nonequilibrium,guioth2020nonadditive}
that in the weak exchange rate limit, the contact dynamics can be conveniently encoded into a coarse-grained exchange rate 
$\varphi(\Delta N_A;\rho_A,\rho_B)$ with $\Delta N_A =N_A'-N_A$ the number of exchanged particles during a single transition, and $\rho_A$ and $\rho_B$ the densities in each system.
In the limit of a large total volume $V_{\rm T}=V_A+V_B$, the joint stationary distribution $P(\rho_A,\rho_B)$ of the number of particles in systems A and B takes the large deviation form
\be \label{eq:large:dev}
P(\rho_A,\rho_B)  \asymp e^{-V_{\rm T}\, I(\rho_A,\rho_B)} \,,
\ee
where the symbol $\asymp$ denotes logarithmic equivalence.
The large deviation function $I(\rho_A,\rho_B)$ has been shown \cite{guioth2018large,guioth2019nonequilibrium} to obey the so-called Hamilton-Jacobi equation
\bea \label{eq:HJ}
&& \sum_{\Delta N_A \ne 0} \Big( \varphi(\Delta N_A;\rho_A,\rho_B)\, e^{I'(\rho_A,\rho_B) \Delta N_A}\\ \nonumber
&& \qquad \qquad \qquad \qquad \qquad - \varphi(-\Delta N_A;\rho_A,\rho_B) \Big) =0\,.
\eea
The derivative $I'$ in Eq.~(\ref{eq:HJ}) is defined as
\be
\label{eq:def:I'}
I' \equiv \frac{1}{\gamma} \frac{\mathrm{d}}{\mathrm{d}\rho_A} I\big(\rho_A,\rho_B(\rho_A)\big)
= \frac{1}{\gamma}\frac{\partial I}{\partial \rho_A} - 
\frac{1}{1-\gamma}\frac{\partial I}{\partial \rho_B}
\ee
having taken into account the conservation law
\be
\gamma \rho_A + (1-\gamma) \rho_B=\overline{\rho} \equiv N_{\rm T}/V_{\rm T} \,,
\ee
where $\gamma=V_A/V_{\rm T}$ is a geometric factor.
A situation of specific interest is when the large deviation function is additive, namely
\be
\label{eq:additivity}
I(\rho_A,\rho_B)=\gamma I_A(\rho_A)+ (1-\gamma) I_B(\rho_B)\,.
\ee
In terms of the derivative $I'$, the addivity condition takes the simple form
\be \label{eq:additivity:Iprime}
I'(\rho_A,\rho_B) = I_A'(\rho_A) - I_B'(\rho_B) \,,
\ee
which allows for the definition of a nonequilibrium chemical potential for the systems in contact \addjg{$\mu_{k}^{\rm cont}(\rho_{k}) = I_{k}'(\rho_{k})$ ($k=A,B$)}
\cite{bertin2006def,bertin2007intensive,pradhan2010nonequilibrium,pradhan2011approximate,chatterjee2015zeroth,
guioth2018large,guioth2019nonequilibrium,guioth2019lack}. In the present framework, the validity or not of the additivity condition is a consequence of the contact dynamics, and it is determined by solving Eq.~(\ref{eq:HJ}).
The latter equation can easily be solved in the particular case when the coarse-grained contact dynamics satisfies the macroscopic detailed balance property defined as
\be \label{eq:HJ:DB}
\varphi(\Delta N_A;\rho_A,\rho_B)\, e^{I'(\rho_A,\rho_B) \Delta N_A} - \varphi(-\Delta N_A;\rho_A,\rho_B) = 0 \,,
\ee
for all $\Delta N_A$. It then follows that
\be \label{eq:Iprime}
I'(\rho_A,\rho_B) = \frac{1}{\Delta N_A}
\ln \frac{\varphi(-\Delta N_A;\rho_A,\rho_B)}{\varphi(\Delta N_A;\rho_A,\rho_B)}
\,,
\ee
the resulting expression being independent of $\Delta N_A$.
Macroscopic detailed balance is obeyed for instance when the stochastic exchange dynamics at contact allows only for single particle exchange.
In this framework, additivity is satisfied when the contact dynamics is factorized between the two systems \cite{guioth2018large,guioth2019nonequilibrium}.
When the contact dynamics is not factorized, or when the macroscopic detailed balance relation (\ref{eq:HJ:DB}) is not obeyed, the large deviation function $I(\rho_A,\rho_B)$ is generically non-additive \cite{guioth2020nonadditive}.
This can be shown for instance by a perturbative expansion around a state satisfying detailed balance \cite{guioth2020nonadditive}.
If the coarse-grained transition rate at contact takes the form $\varphi=\varphi_0 + \varepsilon \varphi_1$, with $\varepsilon \ll 1$, and satisfies macroscopic detailed balance only for $\varepsilon=0$, then the large deviation function $I(\rho_A,\rho_B)$ is determined perturbatively as
\be
I(\rho_A,\rho_B) = I_0(\rho_A,\rho_B) + \varepsilon I_1(\rho_A,\rho_B) + \mathcal{O}(\varepsilon^2)\,.
\ee
While the leading contribution $I_0$ is additive if $\varphi_0$ takes a factorized form, the subleading contribution $I_1$ breaks the additivity property, as seen from its expression \cite{guioth2020nonadditive}
\be \label{eq:Iprime:perturb}
I'_1 = \frac{ \sum_{\Delta N_A \ne 0} \varphi_1(\Delta N_A; \rho_A,\rho_B)
\Big( e^{I_0'(\rho_A,\rho_B)\Delta N_A}-1 \Big)}
{\sum_{\Delta N_A \ne 0} \Delta N_A \varphi_0(\Delta N_A; \rho_A,\rho_B)}
\ee
Having determined the large deviation function $I(\rho_A,\rho_B)$, a multiscale analysis in the slow exchange limit shows that the joint distribution of configurations $P(\mathcal{C}_{A},\mathcal{C}_{B})$ is given to leading order in the small exchange rate by \cite{guioth2020nonadditive}
\begin{equation}
   \label{eq:joint:dist:configs}
   P(\mathcal{C}_{A},\mathcal{C}_{B}) \propto P_A(\mathcal{C}_{A}|\rho_{A}V_{A}) \,P_B(\mathcal{C}_{B}|\rho_{B}V_{B}) \, e^{-V_{\rm T} I(\rho_{A},\rho_{B})} 
\end{equation}
where $P_A(\mathcal{C}_{A}|N_A)$ and $P_B(\mathcal{C}_{B}|N_B)$ are the steady-state configuration distributions in systems A and B when isolated, with fixed particle numbers $N_A=\rho_A V_A$ and $N_B=\rho_B V_B$.
The densities $\rho_A$ and $\rho_B$ are related by the conservation law $\gamma\rho_A+(1-\gamma)\rho_B=\bar{\rho}$.

\section{Building a grand-canonical ensemble}
\label{sec:buildingGC}

\subsection{System in contact with a reservoir}

In the following, we consider system A \addjg{as} the system of interest, and system B is a reservoir whose degrees of freedom are integrated over. To emphasize this different role of the two systems, we slightly change notations and drop the subindex A for quantities characterizing system A (which we simply call `the system' in what follows), while we use from now on the subindex $\mR$ for the reservoir (system B).
The reservoir being by definition much larger than the system of interest, we take the limit $V_{\mR} \to \infty$ keeping fixed the volume $V$ of the system of interest, which implies that the ratio $\gamma=V/(V+V_{\mR}) \to 0$.

The joint distribution $P(\mathcal{C},\mathcal{C}_{\mR})$ given in Eq.~(\ref{eq:joint:dist:configs}) can be integrated over $\mathcal{C}_{\mR}$ to give the distribution of configurations $P(\mathcal{C})$ of the system,
\begin{equation}
   \label{eq:dist:GC:gen}
   P(\mathcal{C}) \propto P(\mathcal{C}|\rho V) \, e^{-V_{\rm T} I_{\gamma}(\rho,\rho_{\mR})} 
\end{equation}
with $\rho_{\mR}=(\bar{\rho}-\gamma\rho)/(1-\gamma)$, and where we have emphasized the $\gamma$-dependence of the large deviation function $I_{\gamma}(\rho,\rho_{\mR})$
[see Eq.~(\ref{eq:additivity}) for the explicit dependence of $I_{\gamma}(\rho,\rho_{\mR})$ on $\gamma$ in the additive case].
The proportionality symbol $\propto$ in Eq.~(\ref{eq:dist:GC:gen}) indicates that the normalization factor is not included explicitly.

We now take the limit $\gamma=V/(V+V_{\mR}) \to 0$ in Eq.~(\ref{eq:dist:GC:gen}), by taking the limit $V_{\mR} \to \infty$ at fixed $V$.
The distribution $P(\mathcal{C}|\rho V)$ of the system considered as isolated does not depend on $\gamma$, since we have fixed the volume $V$. In contrast, the large deviation $I_{\gamma}(\rho,\rho_{\mR})$ generically depends on $\gamma$.
Introducing the most probable value $\rho^{\ast}$ such that $I_{\gamma}(\rho^{\ast},\rho_{\mR}(\rho^{\ast}))=0$ (we assume here that $I_{\gamma}(\rho,\rho_{\mR})$ is a convex function of $\rho$, so that $\rho^{\ast}$ is unique),
we can write, using the definition of $I'$ given in Eq.~\eqref{eq:def:I'},
\begin{equation}
\label{eq:VI:int:Iprime}
V_{\rm T} I_{\gamma}(\rho,\rho_{\mR}) = V \int_{\rho^{\ast}}^{\rho}  I'\left(\rho_1,\frac{\bar{\rho}-\gamma \rho_1}{1-\gamma}\right)\, \mathrm{d}\rho_1 \, .
\end{equation}
It is important to note at this stage that the derivative $I'(\rho,\rho_{\mR})$ of the large deviation function does not depend on $\gamma$ when considered as a function of two independent arguments $\rho$ and $\rho_{\mR}$, because the Hamilton-Jacobi equation \eqref{eq:HJ} does not depend explicitly on $\gamma$.
This is why we write it $I'(\rho,\rho_{\mR})$ instead of $I'_{\gamma}(\rho,\rho_{\mR})$.
It is only when explicitly considering the conservation law $\rho_{\mR}=(\bar{\rho}-\gamma\rho)/(1-\gamma)$ that $\gamma$ comes into play.
Taking the limit $\gamma \to 0$ in Eq.~(\ref{eq:VI:int:Iprime}), we have that $\rho_{\mR} \to \bar{\rho}$.
Since in this limit, $\bar{\rho}$ is also the average density of the reservoir, we will use in the following the notation
$\bar{\rho}_{\mR}$ instead of $\bar{\rho}$, to avoid possible confusion with the average density of the system.
We thus obtain
\begin{equation} \label{eq:VI:int:Iprime:lim}
  V_{\rm T} I_{\gamma}(\rho,\rho_{\mR}) \xrightarrow[\gamma \to 0]{} V \int_{\rho^{\ast}}^{\rho} I'(\rho_1, \bar{\rho}_{\mR}) \mathrm{d}\rho_1 \equiv V J(\rho\vert \bar{\rho}_{\mR})\,,
\end{equation}
where the identity on the right hand side of Eq.~(\ref{eq:VI:int:Iprime:lim}) defines the new function $J(\rho\vert \bar{\rho}_{\mR})$.
Note that for finite $\gamma$, $\rho^{\ast}$ may depend on $\gamma$ if $\gamma$ is varied while keeping ${\bar \rho}_{\mR}$ fixed. In what follows, we assume that the limit $\gamma \to 0$ has been taken, and we define $\rho^{\ast}$ by the relation
\be
J'(\rho^\ast \vert \bar{\rho}_{\mR}) = I'(\rho^\ast, \bar{\rho}_{\mR}) = 0, 
\ee
where $J'(\rho\vert \bar{\rho}_{\mR})$ is the derivative of $J(\rho\vert \bar{\rho}_{\mR})$ with respect to $\rho$.
We can now rewrite Eq.~(\ref{eq:dist:GC:gen}) more explicitly in the limit $\gamma \to 0$ as
\begin{equation}
   \label{eq:dist:GC:gen:gammato0}
   P(\mathcal{C}) \propto P(\mathcal{C}|\rho V) \, e^{-V J(\rho\vert \bar{\rho}_{\mR})} \,.
\end{equation}
Eq.~(\ref{eq:dist:GC:gen:gammato0}) is the most general form of the grand-canonical distribution in the present nonequilibrium setting. We will see below how more explicit forms of the grand-canonical distribution can be obtained under additional assumptions, allowing us to emphasize the similarities and differences with the equilibrium form.
Yet, we first briefly discuss the important issue of the equivalence of the canonical and grandcanonical ensembles.


\subsection{Equivalence of ensembles}

At equilibrium, this equivalence (more precisely the \emph{macrostate} equivalence as defined in \cite{touchette2015equivalence}) is well-known to hold under broad assumptions \cite{touchette2009large,touchette2015equivalence}.
In the present nonequilibrium context, the grand canonical ensemble is formally described by Eq.~\eqref{eq:dist:GC:gen:gammato0}, which allows one to show the equivalence of ensembles in a straightforward way. 
Thanks to the large deviation form of the density distribution in Eq.~(\ref{eq:dist:GC:gen:gammato0}), the grand canonical average of any generic observable $O(\config)$ converges to the same limit as the canonical average in the limit $V \to \infty$, namely
\begin{equation}
  \label{eq:ensemble:equiv}
\lim_{V\to\infty}  \left\langle O(\config) \right\rangle_{\rm GC} 
= \lim_{V\to\infty} \left\langle O \right\rangle_{\rho^{\ast}}
\end{equation}
with $\langle \cdot{} \rangle_{\rho}$ the canonical average for a fixed density $\rho$. The density $\rho^{\ast}$ is the one at which $I'$ is vanishing. Similarly to what happens at equilibrium, equivalence of ensembles no longer holds if the large deviation function $J(\rho\vert \bar{\rho}_{\mR})$ has two or several minima \addjg{(in the presence of phase transition)}.

\subsection{Grand-canonical distribution}
\label{sec:GCdist}

Formally speaking, the grand-canonical distribution is completely prescribed by Eq.~\eqref{eq:dist:GC:gen}. Nonetheless, we will see in the sequel that it appears insightful 
to detail further the expression \eqref{eq:dist:GC:gen:gammato0} when one can define a chemical potential associated with the system taken as isolated.
This will allow us to compare more clearly the nonequilibrium grand-canonical distribution \eqref{eq:dist:GC:gen:gammato0} to its equilibrium counterpart, and to emphasize similarities and differences.

Guided by equilibrium knowledge as well as by the form of nonequilibrium exactly solvable models
\cite{derrida1998asep,evans2004factorized,evans2006factorized,evans2005nonequilibrium,guioth2017mass}, we assume that
\be\label{eq:cano_dist}
P(\mathcal{C}|\rho V)= \frac{F(\mathcal{C})}{Z(\rho, V)}
\ee
for a configuration $\mathcal{C}$ with $N=\rho V$ particles, and $P(\mathcal{C}|\rho V)=0$ otherwise.
The quantity $F(\mathcal{C})$ is the probability weight of configuration $\mathcal{C}$; at equilibrium it would correspond to the Boltzmann-Gibbs factor $e^{-\beta E(\mathcal{C})}$.
$Z(\rho, V)$ is a normalization constant, similar to the partition function at equilibrium,
\be
Z(\rho, V) = \sum_{\mathcal{C}|\rho} F(\config)
\ee
where the sum runs over configurations $\mathcal{C}$ having a given density $\rho$, in a system of fixed volume $V$.
We further assume that the nonequilibrium partition function $Z(\rho,V)$ behaves asymptotically for large volume $V$ as
\be \label{eq:cano_part_func}
Z(\rho,V) \asymp e^{-V\psi(\rho)} \,.
\ee
The rate function $\psi(\rho)$ may be thought of as an effective nonequilibrium free energy density.
Following \cite{bertin2006def,bertin2007intensive}, the nonequilibrium chemical potential $\mu^{\rm iso}(\rho)$ of an isolated system is defined as the derivative of the effective nonequilibrium free energy density,
$\mu^{\rm iso}(\rho)=\psi'(\rho)$. This definition relies on a partition of the isolated systems into virtual subsystems, and ensures equality of the chemical potential $\mu^{\rm iso}(\rho)$ between subsystems.
We may thus write
\be
\psi(\rho) = \psi(\rho^\ast) + \int_{\rho^\ast}^{\rho} \mu^{\rm iso}(\rho_1)\, \mathrm{d}\rho_1 \,.
\ee
From \addjg{\eqref{eq:dist:GC:gen:gammato0}, \eqref{eq:cano_dist} and  \eqref{eq:cano_part_func}}, we thus eventually find for the nonequilibrium grandcanonical distribution of the configuration $\mathcal{C}$,
\begin{equation}
  \label{eq:lambda:def}
  P_{\rm GC}(\mathcal{C}) = \frac{F(\mathcal{C})}{Z_{\rm GC}(V)} \; e^{V\lambda(\rho)}
\end{equation}
with $\rho=\rho(\mathcal{C})$ the density associated with configuration $\mathcal{C}$,
and $Z_{\rm GC}(V)$ a normalization constant.
\addjg{The function $\lambda(\rho)$ introduced in Eq.~(\ref{eq:lambda:def}) is not uniquely defined and can be shifted by a constant, depending on $Z_{\rm GC}(V)$. In order to be consistent with equilibrium, a natural choice is}
\be
\lambda(\rho) = \psi(\rho) - \psi(\rho^\ast) - J(\rho\vert\bar{\rho}_{\mR})\,,
\ee
and may be rewritten as
\be   \label{eq:lambda:def2}
\lambda(\rho) =  \int_{\rho^{\ast}}^{\rho} [ \mu^{\rm iso}(\rho_1)
- I'(\rho_1,\bar{\rho}_{\mR})]\, \mathrm{d}\rho_1 \,.
\ee
Note that the normalization constant $Z_{\rm GC}(V)$ reads for large $V$
\be \label{eq:ZGC}
Z_{\rm GC}(V) \asymp e^{V\lambda(\rho^\ast)} \,.
\ee
We now discuss the properties of the function $\lambda(\rho)$.
If the derivative $I'$ of the large deviation function is additive, as defined by Eq.~(\ref{eq:additivity:Iprime}), $\lambda(\rho)$ is given by
  \begin{equation}
    \label{eq:lambda:additive:rho}
       \lambda(\rho)=\mu_{\mR}^{\rm cont}\, (\rho-\rho^{\ast})
       - \int_{\rho^{\ast}}^{\rho} \eta(\rho_1)\, \mathrm{d}\rho_1 \;,
  \end{equation}
with $\mu_{\mR}^{\rm cont}\equiv \mu_{\mR}^{\rm cont}(\bar{\rho}_{\mR})$, and
where 
\be \label{eq:def:eta}
\eta(\rho) = \mu^{\rm cont}(\rho)-\mu^{\rm iso}(\rho)
\ee
is the difference between the chemical potentials of system in contact with the reservoir, and of the system taken as isolated \cite{guioth2018large,guioth2019nonequilibrium}.
It should be emphasized that in general $\lambda(\rho)$ depends on the contact dynamics between the system and the reservoir, because the function $I'(\rho,\bar{\rho}_{\mathcal{R}})$ itself depends on the contact dynamics \cite{guioth2018large,guioth2019nonequilibrium,guioth2020nonadditive}.

At equilibrium, both chemical potentials are identical, so that $\eta=0$. One thus recovers the standard grandcanonical result  
\be
\lambda_{\rm eq}(\rho)=\mu_{\mR} \, (\rho-\rho^{\ast}) \,,
\ee
with $\mu_{\mR}$ the equilibrium chemical potential of the reservoir.
By contrast, in out-of-equilibrium situations, the chemical potentials $\mu^{\rm cont}(\rho)$ and $\mu^{\rm iso}(\rho)$ take different values, because the chemical potentials $\mu^{\rm cont}(\rho)$ depends on the specific contact dynamics between the two systems in contact \cite{guioth2018large,guioth2019nonequilibrium,guioth2019lack}.
It follows that the quantity $\eta(\rho)$ is nonzero in this case.
In the specific case when $\eta(\rho)$ does not depend on $\rho$, the function $\lambda(\rho)$ remains linear, 
\be
\lambda(\rho)=(\mu_{\mR}^{\rm cont}-\eta) \, (\rho-\rho^{\ast}) \,,
\ee
and one may thus define a nonequilibrium grandcanonical chemical potential
\be \label{eq:muGC}
\mu_{\rm GC} = \mu_{\mR}^{\rm cont}-\eta = \mu_{\mR}^{\rm iso}+\eta_{\mR} -\eta \,.
\ee
where $\eta_{\mR}$ is defined in a similar way as $\eta$ in Eq.~(\ref{eq:def:eta}).
This is expected to happen only for non-interacting particles, as illustrated below in Sec.~\ref{sec:ABP} for a gas of active Brownian particles put in contact with a reservoir through a high potential energy barrier \cite{guioth2019lack}.
In addition, thanks to the equality of the chemical potentials of systems in contact,
$\mu^{\rm cont}(\rho^{\ast})=\mu_{\mR}^{\rm cont}(\bar{\rho}_{\mathcal{R}})$, one finds using the definition (\ref{eq:def:eta}) of $\eta$ that $\mu_{\rm GC} = \mu^{\rm iso}(\rho^{\ast})$. As a result, $\mu_{\rm GC}$ only depends in this case on the intrinsic properties of the system, and not on the contact dynamics with the reservoir. The situation is thus in this respect very similar to equilibrium. Yet, it also differs from equilibrium in the sense that, as seen from Eq.~(\ref{eq:muGC}), the grand-canonical chemical potential $\mu_{\rm GC}$ is not the chemical potential of the reservoir (the latter being $\mu_{\mR}^{\rm cont}$), but $\mu_{\rm GC}$ is rather a parameter characterizing the grand-canonical distribution.

On the other side, if $\eta(\rho)$ is not a constant as a function of $\rho$, Eq.~(\ref{eq:lambda:additive:rho}) shows that $\lambda(\rho)$ is a non-linear function of $\rho$, at odds with equilibrium.
Quite importantly, $\lambda(\rho)$ keeps in this case its dependence on the contact dynamics.

Turning to the case when $I'$ is not additive, one also sees from Eq.~(\ref{eq:lambda:def2})
that $\lambda(\rho)$ is generically non-linear, again precluding the definition of a grandcanonical chemical potential. The case of a nonadditive $I'$ is however difficult to analyze beyond the general expression (\ref{eq:lambda:def2}) without going into explicit examples. This will be the topic of Sec.~\ref{sec:examples}.

Hence whether $I'$ is additive or not, $\lambda(\rho)$ is expected to be a non-linear function of $\rho$. The only exception, as described above, is when $I'$ is additive and $\eta(\rho)$ is independent of $\rho$.
An important consequence of the non-linear behavior of $\lambda(\rho)$ is that the reservoir is no longer characterized by a chemical potential. Hence the usual grandcanonical thermodynamic structure is a priori lost.
For instance, at equilibrium, the variance of the particle number can be expressed as the derivative of the average density with respect to the grand-canonocal chemical potential (which at equilibrium is the chemical potential of the reservoir). This relation cannot be transposed to the nonequilibrium case when, as discussed above, 
a grand-canonical chemical potential can no longer be defined.
However, we will see below how this thermodynamic relation can be restored by applying an external potential, under certain assumptions on the dynamics.

\subsection{Fluctuation-response relation}

We now discuss how to define a fluctuation-response relation when a grand-canonical chemical potential cannot be defined. The idea is to \addjg{probe the response of the coupled systems in contact by applying} an external potential difference $U$ between the system and the reservoir. \addjg{Under the assumptions that macroscopic detailed balance is obeyed and that the perturbation is linear in the thermodynamic forces \cite{guioth2020nonadditive}}, the large deviation function $J(\rho\vert \bar{\rho}_{\mR})$ is modified into
\be
J(\rho\vert \bar{\rho}_{\mR}, U) = J(\rho\vert \bar{\rho}_{\mR}) + \beta U \rho\,,
\ee
where $\beta$ is the inverse temperature \addjg{of the heat bath}. 
The most probable density $\rho^{\ast}_U$ in the presence of the potential difference $U$ thus satisfies
\be
\frac{\partial J}{\partial \rho}(\rho^{\ast}_U\vert \bar{\rho}_{\mR}, U) = J'(\rho^{\ast}_U\vert \bar{\rho}_{\mR}) + \beta U=0 \, ,
\ee
with $J' = \partial J / \partial \rho$.
It follows that 
\be \label{eq:response}
{\left.\frac{\mathrm{d} \rho^{\ast}_U}{\mathrm{d} U} \right| }_{U=0} = -\frac{\beta}{J''(\rho^{\ast}\vert \bar{\rho}_{\mR})}\,,
\ee
with $\rho^{\ast} = \rho^{\ast}_{U=0}$.
On the other side, the density distribution $P(\rho) \asymp e^{-VJ(\rho\vert\bar{\rho}_{\mR})}$ can be approximated close to $\rho^{\ast}$ by a Gaussian distribution, by expanding $J(\rho\vert\bar{\rho}_{\mR})$ to second order in
$\rho-\bar{\rho}_{\mR}$. This Gaussian approximation yields for the variance of the particle number
\be \label{eq:variance}
\textrm{Var}(N) = V^2 \textrm{Var}(\rho) = \frac{V}{J''(\rho^{\ast}\vert \bar{\rho}_{\mR})}\,.
\ee
Combining Eqs.~(\ref{eq:response}) and (\ref{eq:variance}), we obtain the fluctuation-response relation
\be \label{eq:fluct:response}
\frac{\mathrm{d} \rho^{\ast}_U}{\mathrm{d} U}_{\vert U=0} = -\frac{\beta}{V}\, \textrm{Var}(N)\,.
\ee
This fluctuation-response relation generalizes the equilibrium one to the nonequilibrium case when no chemical potential can be defined for the reservoir. Note that when a nonequilibrium chemical potential $\mu_{\rm GC}$ can be defined as in Eq.~(\ref{eq:muGC}), the fluctuation-response relation takes an equilibrium form,
\be \label{eq:fluct:response:eqlike}
\frac{\mathrm{d} \rho^{\ast}}{\mathrm{d} \mu_{\rm GC}} = \frac{\beta}{V}\, \textrm{Var}(N)\,.
\ee
In the following, we discuss two explicit models where the grand-canonical distribution can be determined.

\section{Application to specific models}
\label{sec:examples}

\subsection{Gas of noninteracting active particles}
\label{sec:ABP}

We start by the simple example of a gas of noninteracting active particles, considering either Active Brownian Particles (ABP) or Run-and-Tumble Particles (RTP) \cite{cates2013when}, which model experiments on active colloids \cite{Palacci2010,Theurkauff2012,Bechinger2013,palacci2013living,Bartolo2013,Bechinger2016}, self-propelled grains \cite{Dauchot,Ramaswamy-exp,Kudrolli} or bacteria \cite{Staruss,Aranson}.
As discussed in \cite{guioth2019lack}, such a gas can be split into two compartments in contact through the use of a potential energy barrier $U(\mathbf{r})$ (see also \cite{rodenburg2018ratchet,woillez2020nonlocal}).
We take the second compartment to be much larger than the first one, so that it plays the role of a reservoir $\mR$ of particles.

In two dimensions, the overdamped dynamics of the position $\mathbf{r}=(x,y)$ of an active particle reads
\be
\dot{\mathbf{r}} = v_0\, \mathbf{e}(\theta) - \kappa \nabla U
\ee
where $v_0$ is self-propulsion speed, $\theta$ the polarity angle of the particle, along which the self-propulsion force is applied, and $\kappa$ is a mobility coefficient.
Different models for the dynamics of the angle can be used, for instance a diffusive dynamics $\dot{\theta}=\xi(t)$ for ABPs, with $\xi(t)$ a white noise of diffusion coefficient $D_{\rm r}$, or a jump dynamics with rate $\alpha$ for RTPs \cite{cates2013when}.
It is convenient to assume that the potential $U(\mathbf{r})$ depends only on $x$, and is invariant along the $y$-direction.
We set the origin $x=0$ at the maximum of the potential barrier, and define the system of interest to be on the positive part of the $x$-axis, while the reservoir is on the negative part of the axis.
Taking the limit of a fast angular dynamics, corresponding to a large $D_{\rm r}$ or a large $\alpha$, one finds that the large deviation function $I(\rho_A,\rho_B)$ is additive, and one can compute the chemical potential of the two systems in contact~\cite{guioth2019lack}
\be
\mu^{\rm cont}(\rho) = \ln \rho + \eta \,, \quad 
\mu_{\mR}^{\rm cont}(\rho_{\mR}) = \ln \rho_{\mR} + \eta_{\mR}
\ee
where $\mu^{\rm iso}(\rho)=\ln \rho$ is the chemical potential of the isolated ideal gas, whereas $\eta$ and $\eta_{\mR}$ are the nonequilibrium corrections due to the contact. To leading order in $1/D_{\rm r}$ or $1/\alpha$, the correction $\eta_k$ is given by
\be
\eta = \eta_0 \int_{x^{\ast}}^0 dx \, [U'(x)]^3  \,, \quad 
\eta_{\mR} = \eta_0 \int_{x_{\mR}^{\ast}}^0 dx \, [U'(x)]^3 \,,
\ee
with $\eta_0=\frac{7}{2}\kappa^3 D_{\rm r}^2 /v_0^4$ for active Brownian particles and
$\eta_0=2\kappa^3 \alpha^2 /v_0^4$ for run-and-tumble particles
(note that the large $D_{\rm r}$ or $\alpha$ limit is taken by fixing the effective positional diffusion coefficient
$D_{\rm ABP}=v_0^2/2D_{\rm r}$ or $D_{\rm RTP}=v_0^2/2\alpha$.)
Here, $x^{\ast}>0$ and $x_{\mR}^{\ast}<0$ are arbitrary points in the bulk of the system and of the reservoir respectively.
The parameters $\eta$ and $\eta_{\mR}$ depend explicitly on the shape of the barrier, that is, on the details of the contact dynamics. 
However, $\eta$ and $\eta_{\mR}$ do not depend on density, which results from the assumption of noninteracting particles.
Following the results of Sec.~\ref{sec:GCdist}, it is possible to define in this case a nonequilibrium grandcanonical chemical potential $\mu_{\rm GC}$ for the system in contact with a reservoir of particles.
This grandcanonical chemical potential $\mu_{\rm GC}$ is equal to the chemical potential potential $\ln \rho^{\ast}$ of the ideal gas.
Hence in this case, in spite of the non-trivial corrections $\eta$ and $\eta_{\mR}$ coming from the detailed shape of the potential energy barrier in the evaluation of the chemical potential $\mu^{\rm cont}(\rho)$ and $\mu_{\mR}^{\rm cont}(\rho_{\mR})$, the grand-canonical distribution takes the same form as at equilibrium
(except that, as mentioned above, $\mu_{\rm GC}$ is not the chemical potential of the reservoir).

\subsection{Mass transport model on a lattice}

We now turn to an example of model where the large deviation function is not additive, which can be realized for instance in interacting lattice particle models. Such lattice models are useful benchmarks to test nonequilibrium concepts, as exemplified by the paradigmatic Katz-Lebowitz-Spohn (KLS) model \cite{katz1984nonequilibrium}.
Following \cite{guioth2019nonequilibrium,guioth2020nonadditive}, we consider here the mass transport model introduced in \cite{guioth2017mass} as a driven generalization of the model defined in \cite{bertin2005subdiffusion},
focusing here on the continuous mass version of the model.
Continuous mass transport models have been shown to provide a convenient framework to study for instance condensation transitions \cite{Evans2006}. The present model has the advantage that the probability distribution of configurations depends on the drive, at variance with more standard mass transport models \cite{evans2004factorized,Evans2006}.

The model is defined as follows: 
 on each site $i=1,\dots,2L$ of a one-dimensional lattice with periodic boundary conditions ($2L+1\equiv 1$), a continuous mass $m_i \ge 0$ is defined.
The dynamics of $m_i$ proceeds by sublattice parallel updates, where one of the two partitions of links
$\mathcal{P}_1=\{(2k-1,2k), k=1,\dots,L\}$ and $\mathcal{P}_2=\{(2k,2k+1), k=1,\dots,L\}$ are randomly chosen at each step with equal probabilities.
Having chosen a partition $\mathcal{P}_a$ ($a=1$ or $2$), masses $(m_i,m_{i+1})$ on each link $(i,i+1)$ of the chosen partition are updated in parallel to $(m'_i,m'_{i+1})$ according to the probability
\be
\mathcal{T}(m'_i\vert m_i,m_{i+1}) = \frac{e^{-\eps(m_i+m_{i+1})+\frac{1}{2} f(m'_{i+1}-m'_i)}}{Q(m_i+m_{i+1})}
\ee
and where $m'_{i+1}=m_i+m_{i+1}-m'_i$ is given by the local mass conservation.
The parameter $\eps$ can be interpreted as an effective energy per particle, whereas $f$ is a driving force breaking microscopic detailed balance.
The function $Q$ ensure the normalization condition 
$\int dm'_i \, \mathcal{T}(m'_i\vert m_i,m_{i+1}) =1$.

We now consider two copies of the model, possibly with different parameters, and put them in weak contact
by allowing them to exchange particles at a slow rate.
We choose the so-called Sasa-Tasaki contact dynamics \cite{sasa2006steady}, which depends only on the configuration of the system from which mass is transferred.
We consider that the second model is much larger than the first one, and plays the role of a reservoir of mass.
The corresponding coarse-grained transition rate at contact has been evaluated in \cite{guioth2020nonadditive} and reads
\begin{widetext}
\be
\varphi(\Delta m;\rho_A,\rho_B) = 
\begin{cases}
        e^{\mu^{\mathrm{iso}}_{\mR}(\rho_{\mR}) \Delta m} \left[ \cosh(f_{_{\mR}}\Delta m) + \nu_{\mR}(\rho_{\mR}) f_{\mR}\sinh(f_{\mR} \Delta m) \right]  \quad &\text{for} \; \Delta m > 0 \\
        e^{\mu^{\mathrm{iso}}(\rho) |\Delta m|} \left[ \cosh(f \Delta m) + \nu(\rho) f \sinh(f |\Delta m|) \right]  \quad  &\text{for} \; \Delta m <0 
\end{cases}
\ee
\end{widetext}
with $\nu(\rho)=[\eps-\mu^{\mathrm{iso}}(\rho)]^{-1}$ and
$\nu_{\mR}(\rho)=[\eps_{\mR}-\mu^{\mathrm{iso}}_{\mR}(\rho_{\mR})]^{-1}$.
In this model, macroscopic detailed balance at contact does not hold, and the solution of the Hamilton-Jacobi equation \ref{eq:HJ} has to be evaluated perturbatively in terms of the driving forces $f_A$ and $f_B$, using the perturbative solution given in Eq.~\ref{eq:Iprime:perturb}.
To leading order, one obtains for the derivative of the large deviation function $I'(\rho|\bar{\rho}_{\mR})$ \cite{guioth2020nonadditive}, 
\bea
I'(\rho|\bar{\rho}_{\mR}) &=& \mu^{\mathrm{eq}}(\rho) - \mu^{\mathrm{eq}}(\bar{\rho}_{\mR}) \\ \nonumber
   & \quad & + (f^{2}-f_{\mR}^{2})\frac{\mu^{\mathrm{eq}}(\rho)^{2} + \mu^{\mathrm{eq}}\bar{\mu}_{\mR}^{\mathrm{eq}} + (\bar{\mu}_{\mR}^{\mathrm{eq}})^{2}}{\mu^{\mathrm{eq}}(\rho)^{2} + 2\mu^{\mathrm{eq}}(\rho) \bar{\mu}_{\mR}^{\mathrm{eq}} + (\bar{\mu}_{\mR}^{\mathrm{eq}})^{2}}\,,
\eea
neglecting terms of higher order in $f$ and $f_{\mR}$.
The notation $\bar{\mu}_{\mR}^{\mathrm{eq}} \equiv \mu_{\mR}^{\mathrm{eq}}(\bar{\rho}_{\mR})$ has been introduced.
The equilibrium chemical potential $\mu^{\mathrm{eq}}(\rho)$ is given by
\be
\mu^{\mathrm{eq}}(\rho) = \varepsilon - \frac{1}{\rho} \,.
\ee
To evaluate $\lambda(\rho)$ from Eq.~(\ref{eq:lambda:def2}), we need to know the chemical potential $\mu^{\rm iso}(\rho)$ of the isolated system.
To quadratic order in the drives, $\mu^{\rm iso}$ is given by $\mu^{\rm iso}(\rho)=\mu^{\mathrm{eq}}(\rho)-\frac{1}{4}f^2 \rho$ \cite{guioth2017mass}.
We then find for $\lambda(\rho)$, again to quadratic order in the drive, 
\bea \label{eq:lambda:rho:mtm}
\lambda(\rho) &=& \bar{\mu}_{\mR}^{\mathrm{eq}} (\rho-\rho^\ast) - \frac{1}{8}f^2 (\rho^2-\rho^{\ast 2}) 
- (f^{2}-f_{\mR}^{2}) \times \\ \nonumber
&\times& \int_{\rho^\ast}^{\rho} d\rho_1 \, \frac{(\eps\rho_1-1)^2+\bar{\mu}_{\mR}^{\mathrm{eq}} \rho_1 (\eps\rho_1-1) + (\bar{\mu}_{\mR}^{\mathrm{eq}} \rho_1)^2}{(\eps\rho_1-1)^2+2\bar{\mu}_{\mR}^{\mathrm{eq}} \rho_1 (\eps\rho_1-1) + (\bar{\mu}_{\mR}^{\mathrm{eq}} \rho_1)^2} \,,
\eea
showing explicitly on this example how $\lambda(\rho)$ becomes nonlinear when the drive is switched on.
Several comments are in order here.
First, the most probable density $\rho^\ast$ is also a function of the drive, and it should also be determined
perturbatively in the drive. Yet, taking into account this correction to $\rho^\ast$ in the terms that are quadratic in the drive in Eq.~(\ref{eq:lambda:rho:mtm}) would lead to higher order terms, and $\rho^\ast$ can thus be replaced by its equilibrium value here.
Second, it is interesting to note that $\lambda(\rho)$ remains nonlinear when only the reservoir is subjected to a drive
($f_{\mR} \ne 0$), and the system itself is undriven ($f=0$).
Finally, note that as expected, $\lambda(\rho)$ is an even function of the drives.

\section{Conclusion}

The definition of a grand-canonical ensemble is a difficult task out of equilibrium. Here, we have followed a physically motivated path by explicitly considering that the system of interest is in contact with a much larger system playing the role of a reservoir, which may be either at or out of equilibrium.
Our approach is based on the formalism developed in \cite{guioth2018large,guioth2019nonequilibrium} to define nonequilibrium chemical potentials for systems in contact, through the study of the large deviation function of the densities of the two systems when this large deviation function satisfies an additivity property.
We have also shown that the nonequilibrium grandcanonical ensemble can also be defined when the large deviation function is non-additive \cite{guioth2020nonadditive}.
With the exception of the case of noninteracting particles, for which results similar to the equilibrium grandcanonical ensemble are recovered, the nonequilibrium grandcanonical distribution of microscopic configurations includes an exponential factor $\exp[V \, \lambda(\rho)]$ in place of the standard simple exponential $\exp(V\mu \rho)$. The function $\lambda(\rho)$ is generically non-linear, and depends on the details of the contact dynamics between the reservoir and the system. In this situation, no grandcanonical chemical potential can be defined, although the nonequilibrium grand-canonical ensemble remains well-defined. In a sense, the reservoir is no longer characterized by a single quantity (the chemical potential), but rather by a full function $\lambda(\rho)$ [or perhaps more precisely its derivative $\lambda'(\rho)$] which encodes both the internal properties of the reservoir and that of the contact. Simple thermodynamic relations relating the mean or variance of the number of particles to a derivative of a grand-canonical potential with respect to the chemical potential are thus a priori lost. However they can be recovered for a subclass of contact dynamics provided one introduces a small external potential difference between the system and the reservoir.

It would be of interest to investigate in future work how this grand-canonical ensemble construction is modified when considering a non-vanishing exchange rate with the reservoir. The problem of a finite exchange rate is certainly difficult to address in a general framework, but the assumption that one system is an ideal reservoir with fast relaxation might bring some simplifications to the problem.

\bibliography{biblio_grandcan}

\end{document}